\documentclass[12pt]{article} 
\usepackage{epsfig}

\begin{document} 
\setlength{\parskip}{0.45cm} 
\setlength{\baselineskip}{0.75cm} 
%
%
%
\begin{titlepage} 
\setlength{\parskip}{0.25cm} 
\setlength{\baselineskip}{0.25cm} 
\begin{flushright} 
DO-TH 2000/13\\ 
\vspace{0.2cm} 
hep--ph/0009348\\ 
\vspace{0.2cm} 
September 2000 
\end{flushright} 
\vspace{1.0cm} 
\begin{center} 
\LARGE {\bf Has the QCD RG--Improved Parton Content of Virtual 
Photons been Observed?} \vspace{1.5cm} 
 
\large 
M. Gl\"uck, E.\ Reya, I. Schienbein\\ 
\vspace{1.0cm} 
 
\normalsize 
{\it Institut f\"ur Physik, Universit\"{a}t Dortmund}\\ 
{\it D-44221 Dortmund, Germany} \\ 
 
\vspace{1.5cm} 
\end{center} 
 
\begin{abstract} 
\noindent It is demonstrated that present $e^+e^-$ and DIS ep 
data on the structure of the virtual photon can be understood 
entirely in terms of the standard `naive' quark--parton model box 
approach.  Thus the QCD renormalization group (RG) improved parton 
distributions of virtual photons, in particular their gluonic 
component, have not yet been observed.  The appropriate 
kinematical regions for their future observation are pointed out 
as well as suitable measurements which may demonstrate their 
relevance. 
\end{abstract} 
 
\end{titlepage} 
 
\renewcommand{\theequation}{\arabic{section}.\arabic{equation}} 
\section{Introduction} 
 
Recent measurements and experimental studies of dijet events in 
deep inelastic $ep$ \cite{ref1} and of double--tagged $e^+e^-$ 
\cite{ref2} reactions have indicated a necessity for assigning a 
(QCD resummed) parton content of virtual photons $\gamma(P^2)$ as 
suggested and predicted theoretically [3 -- 9]. 
In particular the DIS dijet production data \cite{ref1} appear to 
imply a sizeable gluon component $g^{\gamma(P^2)}(x,Q^2)$ in the 
derived effective parton density of the virtual photon, where 
$Q^2$ refers to the hadronic scale of the process, $Q\sim p_T^{\rm 
{jet}}$, or to the virtuality of the probe photon $\gamma^*(Q^2)$ 
which probes the virtual target photon $\gamma(P^2)$ in 
$e^+e^-\to e^+e^-X$.  It is the main purpose of this article to 
demonstrate that this is \underline{not} the case and that all 
present data on virtual photons can be explained entirely in 
terms of the conventional QED doubly--virtual box contribution 
$\gamma^*(Q^2)\,\gamma\,(P^2)\to q\bar{q}$ in fixed order 
perturbation theory -- sometimes also referred to as the 
quark--parton model (QPM). 
 
This is of course in contrast to the well known case of a real 
photon $\gamma\equiv\gamma(P^2\equiv 0)$ whose (anti)quark and 
gluon content has been already experimentally established (for 
recent reviews see \cite{ref10,ref11}) which result mainly from  
resummations (inhomogeneous evolutions) of the pointlike mass 
singularities proportional to $\ln\, Q^2/m_q^2$ occurring in the  
box diagram of $\gamma^*(Q^2)\gamma\to q\bar{q}$ for the light 
$q=u,\, d,\, s$ quarks.  This is in contrast to a virtual photon 
target where $\gamma^*(Q^2)\, \gamma\, (P^2)\to q\bar{q}$ does 
\underline{not} give rise to collinear (mass) singularities but 
instead just to finite contributions proportional to $\ln\, 
Q^2/P^2$ which a priori need not be resummed to all orders in QCD. 
 
In Section 2 we shall present the usual QED box contributions 
to the virtual photon structure functions and summarize in the 
Appendix the rather involved exact results in a compact form 
which include all $P^2/Q^2$ as well as $m_q^2/Q^2$ contributions,  
since the latter ones are also important for heavy quark ($c,\, 
b,\, t)$ production.  In Sec.\ 3 we recapitulate briefly how 
these results are resummed in QCD and how (anti)quark and gluon  
distributions in virtual photons are modeled and generated in 
QCD, while Sec.\ 4 contains a comparison with present $e^+e^-$ 
and DIS ep data.  Suggestions of experimental signatures which can 
probe the QCD parton content, in particular the gluon content of 
virtual photons are presented in Sec.\ 5 and our conclusions are 
finally drawn in Section 6. 
 
\section{Virtual Photon Structure Functions and the QED Box Contributions} 
 
The virtual photon structure functions arising in the process 
$e^+(p_1)\,e^-(p_2)\to e^+(p_1\,')\,e^-(p_2\,')\,+\,$ hadrons are 
specified by the kinematical variables 
$q=p_1-p_1\,',\quad 
p=p_2-p_2\,',\quad Q^2=-q^2,\quad P^2=-p^2, \quad y_1=q\cdot 
p/p_1\cdot p$ and $y_2=p\cdot q/p_2\cdot q\, $. 
The Bjorken limit is given by $P^2\ll Q^2$, i.e.\ the virtuality of the target 
photon being small as compared to the one of the probe photon, and 
the corresponding Bjorken variable is $x=Q^2/2p\cdot q$ where 
$0\leq x\leq (1+P^2/Q^2)^{-1}$. The physically measured effective 
structure function in the Bjorken limit is 
\cite{ref12,ref13,ref11} 
\begin{equation} 
\frac{1}{x}\,{F_{\rm{eff}}(x;\, Q^2,y_1;\, P^2,\,y_2)  
= F_{\rm{TT}} + \varepsilon(y_1)F_{\rm{LT}}+ \varepsilon(y_2)F_{\rm{TL}} + 
\varepsilon(y_1) \, \varepsilon(y_2)\, F_{\rm{LL}}\, ,} 
\end{equation} 
where T and L refer to the transverse and longitudinal 
polarization, respectively, of the probe and target photons, and 
$\varepsilon(y_i)$ are the ratios of longitudinal to transverse 
photon fluxes, 
%
\begin{equation} 
\varepsilon(y_i) = 2(1-y_i)/[1+(1-y_i)^2]\, , 
\end{equation} 
and where furthermore  
$F_{ab} = F_{ab}(x,Q^2,P^2)$ with $a=(\rm{L,T}),\quad b=(\rm{L,T})$ .  
In the following we shall consider the kinematical region $y_i\ll 1$ relevant 
for double--tag experiments \cite{ref2,ref14} performed thus far.  Thus  
eq.\ (2.1) reduces to 
%
\begin{equation} 
\frac{1}{x}\, F_{\rm{eff}}(x,Q^2,P^2)\simeq F_{\rm{TT}}+F_{\rm{LT}} 
+F_{\rm{TL}}+F_{\rm{LL}}\,  
\end{equation} 
and ususally one defines \cite{ref11,ref12,ref13} 
%
\begin{eqnarray} 
\frac{1}{x}F_2 & = & F_{\rm{TT}}+F_{\rm{LT}}-\frac{1}{2}\,(F_{\rm{TL}}+F_{\rm{LL}})\nonumber\\ 
 2F_1 & \equiv & 2 F_{\rm{T}} = \bar{\beta}\, ^2(F_{\rm{TT}}-\frac{1}{2}F_{\rm{TL}}) 
\end{eqnarray} 
where $\bar{\beta}\,^2 = 1-4x^2P^2/Q^2$, and $F_{\rm{L}}=\bar{\beta}\,^2F_2-2x\, F_1$. 
(Note that the $F_{ab}$ are normalized with respect to $\frac{1}{x}F_2$, i.e.\ 
$F_{ab}\equiv (Q^2/4\pi^2\alpha)(x\bar{\beta})^{-1}\sigma_{ab}$ with $\sigma_{ab}$ 
denoting the directly measurable cross sections.)  So far, our results are entirely 
general. 
 
We shall furthermore introduce the decomposition 
%
\begin{equation} 
F_{ab} = F_{ab}^{\ell} + F_{ab}^h 
\end{equation} 
with $F\,_{ab}^{\ell(h)}$ denoting the light quark $q = u,\, d,\, s$ (heavy quark  
$h = c,\, b,\, t$) contributions, respectively.  The relevant (QPM) expressions of 
the fully virtual $(P^2\neq 0)$ box for $F_{ab}$ are summarized in the Appendix: 
The light $u,\, d,\, s$ contributions to $F_{ab}^{\ell}$ are obtained from eqs.\ 
(A.1) -- (A.4) by setting $m\equiv m_q = 0\, (\lambda = 0)$ and summing over 
$q = u,\, d,\, s.$  (Note that the box expressions involving a real photon,  
$\gamma^*(Q^2)\, \gamma\, (P^2=0)\to q\bar{q}$, require on the contrary a finite 
regulator mass $m\equiv m_q \neq 0$; here one ususally chooses $m_q$ to be,  
somewhat inconsistently, a constant, i.e.\ $Q^2$--independent effective constituent 
mass, $m_q \simeq 0.3$ GeV.) 
For each heavy quark flavor $h = c,\, b,\, t$ the heavy contribution $F_{ab}^h$ 
in (2.5) is obtained from (A.1) -- (A.4) with $e_q \equiv e_h$ and $m\equiv m_h$. 
Only charm gives a non--negligible contribution for which we choose $m_c = 1.4$ 
GeV throughout. 
 
Finally, it is instructive to recall the asymptotic results of our virtual  
($P^2\neq 0$) box expressions for the light $q = u,\, d,\, s$ quarks derived  
from (A.1) -- (A.4) in the Bjorken limit $P^2/Q^2\ll 1$:   
\begin{eqnarray} 
F_{\rm TT}^{\ell}&  \simeq & 3(\Sigma e_q^4)\, \frac{\alpha}{\pi}  
  \left\{ \left[ x^2+(1-x)^2 \right]\, \ln\frac{Q^2}{P^2x^2} + 4x(1-x) -2 \right\}\nonumber\\ 
F_{\rm LT}^{\ell} & \simeq & F_{TL}^{\ell}\, \simeq\,  3 (\Sigma e_q^4)\,  
   \frac{\alpha}{\pi}\, 4x(1-x)\nonumber\\ 
F_{\rm LL}^{\ell} & \simeq & 0\, ,  
\end{eqnarray} 
i.e.\, using (2.4), 
\begin{equation} 
\frac{1}{x}\, F_{2,\,\rm{box}}^{\ell}(x,Q^2,P^2)\simeq 3(\Sigma e_q^4)\, \frac{\alpha}{\pi} 
  \left\{ \left[ x^2+(1-x)^2\right] \, \ln \frac{Q^2}{P^2x^2} + 6x(1-x)-2 \right\}\, . 
\end{equation} 
In this limit $F_{\rm{eff}}^{\ell}$ in(2.3) reduces to 
\begin{equation} 
 \frac{1}{x}\, F_{\rm{eff}}^{\ell,\,{\rm{box}}}(x,Q^2,P^2)\simeq  
  \frac{1}{x}\, F_{2,\,\rm{box}}^{\ell} + 
   \frac{3}{2}\, F_{\rm{LT}}^{\ell}\, . 
\end{equation} 
Such a relation holds for the heavy quark contribution $F_{\rm eff}^{h,\, {\rm{box}}}$ 
in the Bjorken limit as well, since also $F_{\rm LL}^h$ in (A.4) for $m\equiv m_h\neq 0$ 
becomes vanishingly small for $P^2\ll Q^2$. 
 
The universal process independent part of the pointlike box expressions in (2.6) 
and (2.7) proportional to $\ln\, Q^2/P^2$ may be used to define formally, as in the 
case of a real photon target \cite{ref15}, light (anti)quark distributions in the 
virtual photon $\gamma(P^2)$: 
\begin{equation} 
\frac{1}{x}\, F_{2,\,{\rm box}}^{\ell} (x,Q^2,P^2)|_{\rm univ.} =  
   F_{\rm TT}^{\ell}|_{\rm univ.} \equiv 
     \sum_{q=u,\, d,\, s} e_q^2 \left[ q_{\rm box}^{\gamma(P^2)} (x,Q^2)  
       + \bar{q}\,_{\rm box}^{\gamma(P^2)} (x,Q^2) \right] 
\end{equation} 
with 
\begin{equation} 
q_{\rm box}^{\gamma(P^2)}(x,Q^2) = \bar{q}\,_{\rm box}^{\gamma(P^2)}(x,Q^2)  
  = 3 e_q^2\, \frac{\alpha}{2\pi}\, [x^2+(1-x)^2]\, \ln \frac{Q^2}{P^2}\, . 
\end{equation} 
It should be noted that these naive, i.e.\ not QCD resummed, box expressions 
do not imply a gluon component in the virtual photon,  
$g_{\rm box}^{\gamma(P^2)}(x,Q^2)=0$. 

\setcounter{equation}{0} 
\section{The QCD Parton Content of Virtual Photons} 
Deep inelastic scattering (DIS) involving virtual photons, $\gamma(P^2)$, 
is somewhat problematic since it turns out that the implementation of the  
physical continuity requirement at $P^2=0$ is nontrivial at the 
next--to--leading order (NLO) level due to kinematical discontinuities 
at this point, as exemplified for example by the different expressions 
in eqs.\ (2.7) and (A.6).  A solution to this problem was proposed in 
\cite{ref9} where part of these discontinuities were smoothed out in the  
construction of the photonic parton distributions $f^{\gamma(P^2)}(x,Q^2)$ 
($f= q,\, \bar{q},\, g$ with $q=u,\, d,\, s)$ and where the remaining  
NLO discontinuities, related to the `direct' contributions, were eliminated 
by calculating these contributions as if $P^2=0$.  Thus whenever these 
virtual photons, with their virtuality being here entirely taken care of 
by the `equivalent photon' flux factors \cite{ref9,ref12,ref16}, are 
probed at a scale $Q^2\gg P^2$ they should be considered as {\underline{real}} 
photons which means that cross sections (Wilson coefficient functions) 
of partonic subprocesses involving $\gamma(P^2)$ should be calculated as 
if $P^2=0$.  It should be stressed that this procedure is not a free option 
but a necessary consistency condition for introducing the concept of the 
resolved parton content of the virtual photon as an alternative to a  
non--resummed fixed order perturbative analysis at $P^2\neq 0$ as, for 
example, the QED box results discussed in the previous Section.  This 
consistency requirement is related to the fact that all the resolved 
contributions due to $q^{\gamma(P^2)}(x,Q^2) = \bar{q}\,^{\gamma(P^2)}(x,Q^2)$ 
and $g^{\gamma(P^2)}(x,Q^2)$ are calculated (evoluted) as if these partons 
are massless  
[3--7, 9], i.e.\ employing  
photon splitting functions for real photons, etc., despite of the fact 
that their actual virtuality is given by $P^2\neq 0$. 
 
This rule implies, for example, that the NLO `direct' contribution 
$C_{\gamma(P^2)}(x)$ to $F_2(x,Q^2,P^2)$ has to be the {\underline{same}} 
$C_{\gamma}(x)$ as for {\underline{real}} photons, i.e.\ has to be inferred 
from eq.\ (A.6) of the real (target) photon subprocess $\gamma^*(Q^2)\gamma 
\to q\bar{q}$, and \underline{not} from eq.\ (2.7) which derives from the  
doubly virtual box $\gamma^*(Q^2)\, \gamma (P^2)\to q\bar{q}$  as originally  
proposed \cite{ref3,ref4}.  Thus for the light $u,\, d,\, s$ flavors we have \cite{ref9} 
\begin{equation} 
\frac{1}{x}\, F_2^{\ell}(x,Q^2,P^2) = 2 \sum_{q=u,\, d,\, s} e_q^2 
 \left\{q^{\gamma(P^2)}(x,Q^2) + \frac{\alpha_s(Q^2)}{2\pi}\,  
   \left[ C_q \otimes q^{\gamma(P^2)} + C_g\otimes g^{\gamma(P^2)}\right] \right\} 
\end{equation}  
with the usual (on--shell) Wilson coefficients $C_ {q,\, g}(x)$ as given, for 
example, in \cite{ref9} and the `direct' $C_{\gamma}(x) = [3/(1/2)]C_g(x)$  
contribution has already been absorbed into the definition of $q^{\gamma(P^2)}(x,Q^2)$ 
which thus refer to the DIS$_{\gamma}$ factorization scheme (see, e.g., eq.\ (4)  
in \cite{ref9}).  The heavy quark (predominantly charm) contribution to $F_2$ in 
(2.5) is given by  
\begin{equation} 
  F_2^h(x,Q^2,P^2) = F_{2,\, {\rm{box}}}^h + F_{2,\, g^{\gamma(P^2)}}^h 
\end{equation} 
with the `direct' box contribution given by eq.\ (A.7) and the `resolved' 
contribution by 
\begin{equation} 
F_{2,\,g^{\gamma(P^2)}}^h(x,Q^2) = \int_{z_{\rm{min}}}^1\, \frac{dz}{z}\, 
   zg^{\gamma(P^2)} (z,\mu_{\rm F}^2)\, f_2^{\gamma^*(Q^2)g^{\gamma}\to 
       h\bar{h}} \left(\frac{x}{z},\, Q^2\right) 
\end{equation} 
where $\frac{1}{x}\, f_2^{\gamma^*(Q^2)\gamma\to h\bar{h}}(x,Q^2)$ is given 
by eq.\ (A.7) with $e_h^4\alpha\to e_h^2\alpha_s(\mu_{\rm F}^2)/6,\, 
z_{\rm min} = x(1 + 4m_h^2/Q^2)$ and $\mu_{\rm F}^2\simeq 4m_h^2.$  
Furthermore, since an effectively real photon has no longitudinal components 
($F_{\rm TL},\, F_{\rm LL}$) we have, instead of (2.3) or (2.8), 
\begin{equation} 
F_{\rm eff}(x,Q^2,P^2) = F_2(x,Q^2,P^2)\, . 
\end{equation} 
Finally it should be remarked that the parton distributions $f^{\gamma}(x,Q^2)$ 
of a real photon can be calculated in a parameter--{\underline{free}} way 
\cite{ref9} by employing a coherent superposition of vector mesons, which  
maximally enhances $u$--quark contributions to $F_2^{\gamma}$, for determining the  
hadronic input $f_{\rm had}^{\gamma}(x,Q_0^2)$ at a GRV--like \cite{ref17} input 
scale $Q_0^2\equiv \mu^2\simeq 0.3$ GeV$^2$.  For a virtual photon, however, 
an additional assumption is needed about the $P^2$--dependence of the hadronic 
input distributions of a virtual photon, $f_{\rm had}^{\gamma(P^2)}(x,Q_0^2)$, 
which is commonly {\underline{assumed}} \cite{ref6,ref7,ref9} to be represented 
by a vector--meson--propagator--inspired suppression factor $\eta(P^2)=(1+P^2/ 
m_{\rho}^2)^{-2}$ with $m_{\rho}^2=0.59$ GeV$^2$.  Thus our above consistency 
requirement affords furthermore the following boundary conditions for quarks 
and gluons \cite{ref9}: 
\begin{equation} 
f^{\gamma(P^2)}(x,Q^2=\tilde{P}^2) = f_{\rm had}^{\gamma(P^2)} (x,\, \tilde{P}^2) 
   = \eta(P^2)\, f_{\rm had}^{\gamma}(x,\, \tilde{P}^2) 
\end{equation} 
in LO as well as in NLO of QCD, where $\tilde{P}^2={\rm max} (P^2,\mu^2)$ 
as dictated by continuity in $P^2$ as well as by the fact that the hadronic 
component of $f^{\gamma(P^2)}(x,Q^2)$ is probed at the scale $Q^2=\tilde{P}^2$ 
\cite{ref5,ref6,ref7,ref9} where the pointlike component vanishes by definition. 
The second equality in (3.5) follows from the consistency requirement that 
$C_{\gamma(P^2)}(x)$ is taken to be given by $C_{\gamma}(x)$ and consequently 
the application of the {\underline{same}} $\overline{\rm{MS}}\to$ DIS$_{\gamma}$ 
factorization scheme transformation as for the real photon \cite{ref9}.  Thus  
the resulting perturbatively stable LO and NLO parton densities $f^{\gamma(P^2)} 
(x,Q^2)$ are {\underline{smooth}} in $P^2$ and apply to {\underline{all}}  
$P^2\geq 0$ whenever $\gamma(P^2)$ is probed at scales $Q^2\gg P^2$. 
 
A different approach has been suggested by Schuler and Sj\"ostrand \cite{ref7}. 
Apart from using somewhat different input scales $Q_0$ and parton densities,  
the perturbatively exactly calculable box expressions for $\Lambda^2\ll P^2\ll Q^2$ 
in (2.6) and (2.7) are, together with their LO--QCD $Q^2$--evolutions, extrapolated  
to the case of real photons $P^2=0$ by employing some dispersion--integral--like 
relations.  These link perturbative and non--perturbative contributions and  
allow a smooth limit $P^2\to 0$. (Note, however, that the LO $Q^2$--evolutions 
are performed by using again the splitting functions of real photons and  
on--shell partons.)  Since one works here explicitly with virtual ($P^2\neq 0)$ 
expressions, the longitudinal contributions of the virtual photon target  
should be also taken into account when calculating $F_{\rm eff} = F_2 + \frac{3}{2} 
F_{\rm LT}$ similarly to (2.8), as described for example in \cite{ref6}, which 
is in contrast to our approach in (3.4). 
 
In an alternative approach \cite{ref18} 
one may consider the longitudinal 
component of the virtual photon target $\gamma_{\rm L}(P^2)$ to possess,  
like the transverse component, a universal process independent hadronic content 
obtained radiatively via the standard homogeneous (Altarelli-Parisi)  
$Q^2$--evolution equations with the boundary conditions for the pointlike  
component at $Q^2=P^2$ given by $F_{\rm TL}^{\ell}$ in eq.\ (2.6) for quarks 
together with a vanishing gluonic input in LO.  We have checked that the  
predictions for $F_{\rm eff}(x,Q^2,P^2)$ obtained in this approach differ 
only slightly (typically about 10\% or less) 
 from those of the standard fixed order perturbative approach at 
presently relevant kinematical regions  
$(P^2$ \raisebox{-0.1cm}{$\stackrel{<}{\sim}\,$} $\frac{1}{10}Q^2$, 
$x$ \raisebox{-0.1cm}{$\stackrel{>}{\sim}$} 0.05) 
due to the smallness of $F_{\rm TL}^{\ell}$ relative to $F_{\rm TT}^{\ell}$ 
in (2.6). 
 
As already mentioned, at $Q^2\gg P^2\gg \Lambda^2$, the $\ln Q^2/P^2$ 
terms in (2.6) and (2.7) need {\underline{not}} necessarily be resummed 
in contrast to the situation for the real photon ($P^2=0$) with its 
well known mass singularities $\ln\, Q^2/m_q^2$ in (A.6) which {\underline{afford}} 
the introduction of scale dependent (RG--improved) parton distributions which 
are a priori unknown unless one resorts to some model assumptions about their 
shape at some low resolution scale (see, e.g., \cite{ref9} and the recent 
reviews \cite{ref10,ref11}). 
 
\setcounter{equation}{0} 
\section{Comparison of Theoretical Expectations with Present $e^+e^-$ 
         Virtual Photon Data}  
We shall now turn to a quantitative study of the various QED-box and QCD  
$Q^2$--evoluted structure function expectations for a virtual photon target 
and confront them with all presently available $e^+e^-$ data of PLUTO  
\cite{ref14} and the recent one of LEP-L3 \cite{ref2}.  Despite the limited 
statistics of present data the box predictions for $F_{\rm eff}$ in (2.3) 
shown in figs.\ 1 and 2 appear to be in even better agreement with present 
measurements than the QCD resummed expectations of SaS \cite{ref7} and  
GRS \cite{ref9}.  Typical QCD effects like the increase in the small-x  
region in fig.\ 1, being partly caused by the presence of a finite gluon 
content $g^{\gamma(P^2)}(x,Q^2)$, cannot be delineated with the present poor 
statistics data. 
 
These results clearly demonstrate that the naive QPM predictions derived 
from the doubly--virtual box $\gamma^*(Q^2)\, \gamma(P^2)\to q\bar{q}$ 
fully reproduce all $e^+e^-$ data on the structure of virtual 
photons $\gamma(P^2)$.  In other words, there is {\underline{no}} sign of a 
QCD resummed parton content in virtual photons in present data, in particular 
of a finite gluon content $g^{\gamma(P^2)}(x,Q^2)$ which is absent in the  
`naive' box (QPM) approach. 
 
Characteristic possible signatures for QCD effects which are caused by the 
presence of a finite and dominant gluon component $g^{\gamma(P^2)}$ will be  
discussed in Sec.\, 6. 
 
\setcounter{equation}{0} 
\section{Comparison of Theoretical Expectations with DIS ep Data and Effective 
Quark Distributions of Virtual Photons} 
In oder to extract the parton densities of virtual photons from DIS ep dijet 
data, the H1 collaboration \cite{ref1} has adopted the `single effective 
subprocess approximation' \cite{ref19} which exploits the fact that the 
dominant contributions to the cross section in LO--QCD comes from the $2\to 2$ 
parton--parton hard scattering subprocesses that have similar shapes and thus 
differ mainly by their associated color factors.  Therefore the sum over the 
partonic subprocesses can be replaced by a single effective subprocess cross 
section and effective parton densities for the virtual photon given by 
\begin{equation} 
\tilde{f}\,^{\gamma(P^2)}(x,Q^2) = \sum_{{\rm q=u,\, d,\, s}} 
  \left[ q^{\gamma(P^2)}(x,Q^2)\, + \bar{q}\,^{\gamma(P^2)}(x,Q^2) \right] 
   + \frac{9}{4}\,g^{\gamma(P^2)}(x,Q^2)  
\end{equation} 
with a similar relation for the proton $\tilde{f}\,^p(x,Q^2)$ which is  
assumed to be known.  It should be emphasized that such an effective 
procedure does not hold in NLO where all additional (very different) $2\to 3$ 
subprocesses contribute \cite{ref20}. This NLO analysis affords therefore a 
confrontation with more detailed data on the triple--differential dijet 
cross--section as compared to presently available data \cite{ref1} which 
are not yet sufficient for examining the relative contributions of  
$q^{\gamma(P^2)}(x,Q^2)$ and $g^{\gamma(P^2)}(x,Q^2)$.  In fig.\ 3 we  
compare our LO RG--resummed predictions for $\tilde{f}\,^{\gamma(P^2)}(x,Q^2)$ 
with the naive non--resummed universal (process independent) box expressions 
in (2.10).  Although the fully QCD--resummed results are sizeable and somewhat 
larger in the small $P^2$ region than the universal box expectations, present 
H1 data \cite{ref1} at $Q^2\equiv (p_{\rm T}^{\rm jet})^2 = 85$ GeV$^2$ 
cannot definitely distinguish between these predictions.  It should be furthermore 
noted that the QCD gluon contribution $g^{\gamma(P^2)}(x,Q^2)$ is suppressed 
at the large values of $x$ shown in fig.\ 3.  Therefore present data \cite{ref1} 
cannot discriminate between the finite QCD resummed component $g^{\gamma(P^2)} 
(x,Q^2)$ and the non--resummed $g_{\rm box}^{\gamma(P^2)}(x,Q^2) = 0$. 
 
It is obvious that these two results shown in fig.\ 3 are only appropriate  
for virtualities $P^2\ll Q^2$, typically $P^2=10$ to 20 GeV$^2$ at $Q^2 = 85$ 
GeV$^2$, since ${\cal{O}}(P^2/Q^2)$ contributions are neglected in  
RG--resummations as well as in the definition (2.10).  In order to demonstrate 
the importance of ${\cal{O}}(P^2/Q^2)$ power corrections in the large $P^2$  
region let us define, generalizing the definition (2.9), some effective 
(anti)quark distributions as common via 
\begin{equation} 
\frac{1}{x}\, F_{2,\, {\rm box}}^{\ell}(x,Q^2,P^2) \equiv 
  \sum_{\rm q=u,\, d,\, s} e_q^2 \left[ q_{\rm eff}^{\gamma(P^2)}(x,Q^2) 
      + \bar{q}_{\rm eff}^{\gamma(P^2)}(x,Q^2)\right] 
\end{equation} 
where, of course, $q_{\rm eff}^{\gamma(P^2)}=\bar{q}_{\rm eff}^{\gamma(P^2)}$ 
and the full box expression for $F_{2,\,{\rm box}}^{\ell}$ in (2.4) for light 
quarks is given in (A.1) -- (A.4) with $m\equiv m_q=0$, i.e.\ $\lambda =0$. 
The full box expressions imply again $g_{\rm eff}^{\gamma(P^2)}(x,Q^2)=0$ 
in contrast to the QCD resummed gluon distribution.  The $q_{\rm eff}^{\gamma(P^2)}$ 
introduced in (5.2) is, in contrast to (2.9), of course non--universal.  The 
`effective' results shown in fig.\ 3 clearly demonstrate the importance of 
the ${\cal{O}}(P^2/Q^2)$ terms at larger values of  
$P^2$ \raisebox{-0.1cm}{$\stackrel{<}{\sim}\,$} $Q^2$  
which are not taken into account by the QCD resummations and by the universal 
box expressions in (2.10) also shown in fig.\ 3.  It is interesting that the 
non--universal $q_{\rm eff}^{\gamma(P^2)}$ defined via $F_2$ in (5.2) describes 
the H1--data at large values of $P^2$ in fig.\ 3 remarkably well.  This may be 
accidental and it remains to be seen whether future LO analyses will indicate 
the general relevance of $q_{\rm eff}^{\gamma(P^2)}(x,Q^2)$ in the large $P^2$ 
region. 
 
As we have seen, present DIS dijet data cannot discriminate between the universal 
naive box and QCD--resummed expectations in the theoretically relevant region 
$P^2\ll Q^2$, mainly because these data are insensitive to the gluon content in 
$\gamma(P^2)$ generated by QCD--evolutions which is absent within the naive box 
approach.  Therefore we finally turn to a brief discussion where such typical 
QCD effects may be observed and delineated by future experiments. 
 
\setcounter{equation}{0} 
\section{Possible Signatures for the QCD Parton Content of Virtual Photons} 
Since $e^+e^-$ and DIS ep dijet data cannot, at present, delineate the  
QCD--resummed parton content of a virtual photon, in particular not its gluon 
content, we shall now propose and discuss a few cases where such typical QCD 
effects may be observed and possibly confirmed by future experiments. 
 
Charm production in $e^+e^-\to e^+e^-\, c\bar{c}X$ would be a classical possibility 
to delineate such effects due to a nonvanishing $g^{\gamma(P^2)}(x,Q^2)$.  In  
fig.\ 4 we compare the usual (fixed order) `direct' box contribution to $F_2^c$ 
with the  `resolved' gluon--initiated one in (3.2), as given by (3.3).  The 
`direct' box contribution entirely dominates in the large $x$ region,   
$x$ \raisebox{-0.1cm}{$\stackrel{>}{\sim}$} 0.05, accessible by present experiments 
(cf.\ figs.\ 1 and 2), whereas the typical QCD--resummed `resolved' contribution 
becomes comparable to the  `direct' one and eventually dominates in the small $x$ 
region, $x<0.05$.  Thus a careful measurement of the charm contribution to $F_2$ 
at  
$x$ \raisebox{-0.1cm}{$\stackrel{<}{\sim}$} 0.05 would shed some light on the QCD 
parton (gluon) content of virtual photons, since such a `resolved' contribution 
in fig.\ 4 would be absent within the naive box approach. 
 
The effective parton distribution $\tilde{f}^{\gamma(P^2)}(x,Q^2)$ in (5.1) at 
not too large values of $x$ and $P^2$, as may be extracted in LO from DIS ep dijet 
data, would be another possibility to observe QCD--resummation effects due to a 
nonvanishing gluon component $g^{\gamma(P^2)}(x,Q^2)$.  In fig.\ 5 we show the quark 
and gluon contributions to $\tilde{f}^{\gamma(P^2)}$ in (5.1) separately.  The box 
(anti)quark contributions, which are similar to the QCD--resummed ones, entirely 
dominate over the QCD--resummed gluon contribution in the large $x$ region,  
$x$ \raisebox{-0.1cm}{$\stackrel{>}{\sim}$} 0.4,  
accessible to present experiments (cf.\ fig.\ 3).  Only {\underline{below} $x\simeq 0.3$ 
does the QCD gluon contribution become comparable to the (anti)quark components  
and dominates, as usual, for  
$x$ \raisebox{-0.1cm}{$\stackrel{<}{\sim}$} 0.1. 
It should be remembered that $g_{\rm box}^{\gamma(P^2)}(x,Q^2)=0$.  Furthermore, 
the increase of the RG--resummed   
$q^{\gamma(P^2)}(x,Q^2)$ at small $x$ in fig.\ 5 
is induced by the vector--meson--dominance--like input for the $Q^2$--evolution 
of the `hadronic' component of photon's parton distribution \cite{ref7,ref9} and 
is disregarded in our naive box (QPM) analysis. 
 
Thus a measurement of dijets produced in DIS ep reactions in the not too large 
$x$ region,  
$x$ \raisebox{-0.1cm}{$\stackrel{<}{\sim}$} 0.3, 
would probe the QCD parton content of virtual photons, in particular their gluon 
content which is absent in the naive QPM box approach.  In this region, and at not 
too large photon virtualities  
$P^2$ \raisebox{-0.1cm}{$\stackrel{<}{\sim}$} 5 GeV$^2$ 
shown in fig.\ 5, the  `resolved' gluon--dominated contribution of the virtual 
photon to high $E_{\rm T}$ jet production at scales $Q\equiv E_{\rm T} \simeq 5 
- 10$ GeV exceeds by far the `direct' box--like contribution of a pointlike 
virtual photon \cite{ref21}.      
 
\section{Summary and Conclusions} 
Virtual photons $\gamma(P^2)$, probed at a large scale $Q^2\gg P^2$, may be 
described either by fixed--order perturbation theory, which in lowest order of  
QCD yield the quarks and antiquarks generated by the universal part of the `box' 
diagram, or alternatively by their renormalization group (RG) improved counterparts 
including particularly the gluon distribution $g^{\gamma(P^2)}(x,Q^2)$. 
 
The results in Sections 3 and 4 demonstrate that all presently available $e^+e^-$ 
and DIS ep dijet data can be fully accounted for by the standard doubly--virtual 
QED box diagram and are not yet sensitive to RG resummation effects which are manifest 
only in the presently unexplored low--$x$ region of the parton distributions in 
$\gamma(P^2)$. In fact, as shown in Section 6, these resummation effects start to 
dominate only a $x<0.3$ and may be observed by future measurements at $P^2={\cal{O}} 
(1$ GeV$^2)$ of $\sigma(ep\to e\, jjX)$ or $\sigma(e^+e^- \to e^+e^-\, c\bar{c}X)$  
at high energy 
collisions.  These measurements could finally discriminate between the fixed order 
and RG--improved parton distributions of the virtual photon.  
\newpage 
 
\renewcommand{\theequation}{A.\arabic{equation}} 
\setcounter{equation}{0} 
\noindent {\bf{\Large{Appendix}}} 
 
\noindent The most general QPM box--results for $F_{ab}$ appearing in the structure 
 function relations (2.3) and (2.4) derive from the fully off--shell QED box--diagram  
$\gamma^*(Q^2)\, \gamma (P^2)\to q\bar{q}$ for each quark flavor with charge 
$e_q$, carrying 3 colors and by keeping the quark mass $m$ as well \cite{ref12,ref4}. 
These results can be conveniently written as : 
\begin{eqnarray}  
F_{\rm TT} &=& 3\, e_q^4\, \frac{\alpha}{\pi}\, \theta(\beta^2)\, 
    \frac{1}{\bar{\beta}^5}\,   
 \Biggl\{ \biggl[ 1-2x(1-x)-2\delta(1-\delta) - 4x\delta(x^2+\delta^2)  
    + 8 x^2\delta^2  
\nonumber\\ 
   &&  \left[1+(1-x-\delta)^2\right]+\lambda\bar{\beta}^2(1-x-\delta)^2 -\frac{1}{2}\lambda^2\bar{\beta}^4 
            (1-x-\delta)^2\biggr] \, \ln\, \frac{\beta_+}{\beta_-}  
\nonumber\\ 
&& + \beta\bar{\beta}\, \Biggl[ 4x(1-x)-1+4\delta(1-\delta)-8x\delta (1-x^2-\delta^2) 
\nonumber\\ 
&&   -(4x\delta+\lambda\bar{\beta}^2)(1-x-\delta)^2 - 
            \frac{4x\delta\bar{\beta}^4}{4x\delta + \lambda\,\bar{\beta}^2}\Biggr] 
       \Biggr\}\\\nonumber \\ 
F_{\rm LT} &=& 3\, e_q^4\, \frac{\alpha}{\pi}\, \theta(\beta^2)\, 
 \frac{4}{\bar{\beta}^5}\, 
 (1-x-\delta)\, \Biggl\{ x\biggl[\, - \frac{1}{2}\lambda\,\bar{\beta}^2 
    \Bigl( 1-2\delta(1+x-\delta)\Bigr) 
\nonumber\\ 
 & &  -2\delta \Bigl(-1+2x+2\delta -2x\delta(1+x+\delta) \Bigr) \biggr]\,  
         \ln \frac{\beta_+}{\beta_-} 
\nonumber\\  
& &   +\beta\bar{\beta}\left[ x(1-6\delta +6\delta^2 +2x\delta)  
         + \delta\bar{\beta}^2\, \frac{4x\delta} 
               {4x\delta +\lambda\, \bar{\beta}^2} \right] \Biggr\} 
\\\nonumber \\ 
F_{\rm TL} &=& F_{\rm LT}\, [x\leftrightarrow\delta]\hspace{9.8cm} 
\\\nonumber \\ 
F_{\rm LL} &=& 3\, e_q^4\, \frac{\alpha}{\pi}\, \theta(\beta^2)\, 
  \frac{16}{\bar{\beta}^5}\, \delta x(1-x-\delta)^2 
    \left\{ (1+2x\delta)\,\, \ln \frac{\beta_+}{\beta_-} 
    \, -2\beta\bar{\beta}\, \frac{6x\delta+\lambda\,\bar{\beta}^2} 
     {4x\delta + \lambda\,\bar{\beta}^2}\right\} 
\end{eqnarray} 
\vspace{-0.3cm} 

\noindent where 
 $\delta=xP^2/Q^2$, $\lambda = 4m^2/W^2$ with $W^2=Q^2(1-x-\delta)/x \geq(2m)^2$, 
and  
$\beta^2=1-\lambda$, $\bar{\beta}^2=1-4x\delta$, $\beta_{\pm}=1\pm\beta\bar{\beta}$.  
The relevant asymptotic expressions for the light $q = u,\, d,\, s$ quark 
($m\equiv m_q=0$, i.e., $\lambda=0$) contributions in the Bjorken limit $P^2\ll Q^2$ 
are given in eq.\ (2.6). 
 
For completeness it should be noted that the general virtual box results in 
(A.1) -- (A.4) reduce for $P^2=0\, (\delta=0)$ to the standard box--diagram 
$\gamma^*(Q^2)\,\gamma\to q\bar{q}$ expressions for a {\underline{real} photon 
$\gamma\equiv\gamma(P^2=0)$ :  in the light quark sector where $\lambda\ll 1$, 
i.e.\ $m^2\equiv m_q^2\ll Q^2$, we have  
%
\begin{eqnarray} 
F_{\rm TT}^{\ell} & = & 3\, \Sigma e_q^4\, \frac{\alpha}{\pi}  
    \left\{ [x^2+(1-x)^2] \, \ln\, \frac{Q^2(1-x)}{m_q^2\, 
        x}+ 4x (1-x)-1 \right\}\nonumber\\ 
F_{\rm LT}^{\ell} & \simeq & 3\, \Sigma e_q^4\, \frac{\alpha}{\pi}\  
      4 x (1-x)\nonumber\\ 
F_{\rm TL}^{\ell} & = & F_{\rm LL}^{\ell} = 0\, , 
\end{eqnarray} 
i.e., according to (2.4) 
%
\begin{equation} 
\frac{1}{x}\, F_{2,\, {\rm box}}^{\ell}(x,Q^2) =  
   3\, \Sigma e_q^4\, \frac{\alpha}{\pi} \left\{ \left[ x^2+(1-x)^2\right]\, 
       \ln\, \frac{Q^2(1-x)}{m_q^2\, x}\, + 8 x(1-x)-1 \right\}\, . 
\end{equation} 
The heavy quark contribution becomes 
%
\begin{eqnarray} 
F_{\rm TT}^h & = & 3\, e_h^4\, \frac{\alpha}{\pi}\, \theta(\beta^2) 
     \left\{ \left[ x^2+(1-x)^2+x(1-x)\, \frac{4m_h^2}{Q^2}\, -x^2\,  
       \frac{8m_h^4}{Q^4}\right]\, \ln \frac{1+\beta}{1-\beta}\right.\nonumber\\ 
 & &     \left.  + \beta\left[ 4x(1-x)-1-x(1-x)\,  
           \frac{4m_h^2}{Q^2}\right]\right\}\nonumber\\ 
F_{\rm LT}^h  & = & 3\, e_h^4\, \frac{\alpha}{\pi}\, \theta(\beta^2) 
     \left\{ -x^2\, \frac{8m_h^2}{Q^2}\, \ln \frac{1+\beta}{1-\beta} +\beta 
       \left[ 4x(1-x)\right] \right\}\nonumber\\ 
F_{\rm TL}^h & = & F_{\rm LL}^h = 0\, ,  
\end{eqnarray} 
i.e., according to (2.4), 
%
\begin{eqnarray} 
\frac{1}{x}\, F_{2,\, {\rm box}}^h(x,Q^2) & = & 3\, e_h^4\, \frac{\alpha}{\pi}\, 
   \theta(\beta^2) 
     \left\{ \left[ x^2+(1-x)^2+x(1-3x)\, \frac{4m_h^2}{Q^2}\,\right.\right.\\ 
& & \left.\left. -x^2\, \frac{8m_h^4}{Q^4}\right] \,  
     \ln\, \frac{1+\beta}{1-\beta} + \beta\left[ 8x(1-x)-1-x(1-x)\,  
       \frac{4m_h^2}{Q^2}\right] \right\}\nonumber 
\end{eqnarray} 
\vspace{-0.3cm} 
 
\noindent and $\frac{1}{x}\, F_{\rm L,\, box}^h(x,Q^2) = F_{\rm LT}^h$, which are the 
familiar massive Bethe--Heitler expressions \cite{ref22} relevant for the  
heavy quark contributions to the structure functions of real photons 
(cf.\ \cite{ref9}, for example).


 
\newpage 
 
\noindent{\large{\bf{\underline{Figure Captions}}}} 
\begin{itemize} 
\item[\bf{Fig.\ 1}]  Predictions for $F_{\rm eff}$ as defined in (2.3).  The light 
      $(u,\, d,\, s)$ and heavy (charm) contributions in (2.5) of the `full box'  
      expressions in (A.1) -- (A.4) are calculated as explained in the text below 
      eq.\ (2.5).  The `asymptotic box' results refer to the light quark contributions 
      being given by (2.6) or (2.7) and (2.8).  The QCD resummed NLO expectations of 
      GRS \cite{ref9} for $F_2$ in (3.4) turn out to be similar to the LO ones  
      \cite{ref9}.  Also shown are the LO--resummed results of SaS 1D \cite{ref7} for 
      $F_2$ and $F_{\rm eff} = F_2 +\frac{3}{2}\, F_{\rm LT}$ as discussed in Sec.\ 3.  
      The total 
      charm contribution to the latter two QCD results involves also a `resolved' 
      component, e.g., eqs.\ (3.2) and (3.3), which turns out to be small as compared 
      to the box contribution shown which dominates in the kinematic region  
      considered.  The PLUTO data are taken from \cite{ref14}. 
\item[\bf{Fig.\ 2}]  As in fig.\ 1, but for $Q^2=120$ GeV$^2$ and $P^2=3.7$ GeV$^2$ 
      appropriate for the LEP--L3 data \cite{ref2}. 
\item[\bf{Fig.\ 3}]  Predictions for the effective parton density defined in eq.\ (5.1). 
      The `box' results refer to the universal $q_{\rm box}^{\gamma(P^2)}$ in (2.10), 
      and the `effective' ones to $q_{\rm eff}^{\gamma(P^2)}$ as defined in (5.2) as  
      derived from the full box expressions (A.1) -- (A.4) including all ${\cal{O}}(P^2/Q^2)$ 
      contributions.  
      The LO--QCD predictions of GRS \cite{ref9} are shown by the solid 
      curves which refer to the predictions in the theoretically legitimate region 
      $P^2\ll Q^2$, whereas the dashed curves extend into the kinematic region of larger 
      $P^2$ approaching $Q^2$ where the concept of QCD--resummed parton distributions 
      of virtual photons is not valid anymore. (Note that the results for $x=0.6$ 
      terminate at $P^2\simeq 
      54$ GeV$^2$ due to the kinematic constraint $W^2>0$, with $W^2$ being defined below 
      (A.4), i.e.\ $x<(1+P^2/Q^2)^{-1}$.)   For illustration we also show the 
      effective LO--QCD parton density $\tilde{f}^{\gamma}$ of a real photon $\gamma\equiv 
      \gamma(P^2=0)$ of GRS \cite{ref9} multiplied by the simple $\rho$--pole suppression 
      factor $\eta(P^2)$ in (3.5) which clearly underestimates the H1--data \cite{ref1}. 
\item[\bf{Fig.\ 4}]  Expected charm contributions to $F_2$.  The naive `direct (box)' 
      result refers to $F_{2,\, {\rm box}}^c$ in (3.2) and the LO--QCD  `resolved' 
      prediction is due to $F_{2,\, g^{\gamma (P^2)}}^c$ in (3.2), as explicitly given  
      in (3.3) with $g^{\gamma(P^2)}(x,\, 4m_c^2)$ taken from GRS \cite{ref9}.  This 
      latter `resolved' contribution is absent in the naive box (QPM) approach. 
\item[\bf{Fig.\ 5}]  Predictions for the total light quark $\Sigma^{\gamma(P^2)}\equiv 
      2\Sigma_{\rm{q = u,\, d,\, s}}\,q^{\gamma(P^2)}$ 
      and gluon contributions to the effective parton density in (5.1) at a fixed scale 
      $Q^2=85$ GeV$^2$ and two fixed virtualities $P^2=1$ and 5 GeV$^2$.  The naive box 
      results refer to the universal $q_{\rm box}^{\gamma(P^2)}$ defined in (2.10), and 
      to $q_{\rm eff}^{\gamma(P^2)}$ defined in (5.2).  The LO--QCD RG--resummed predictions 
      are denoted by $\Sigma^{\gamma(P^2)}$ and $g^{\gamma(P^2)}$ according to GRS \cite{ref9}. 
      The latter gluon contribution is absent in the naive box (QPM) approach. 
\end{itemize} 

\newpage
\pagestyle{empty}
\begin{figure}
\centering
\vspace*{-1cm}
\epsfig{figure=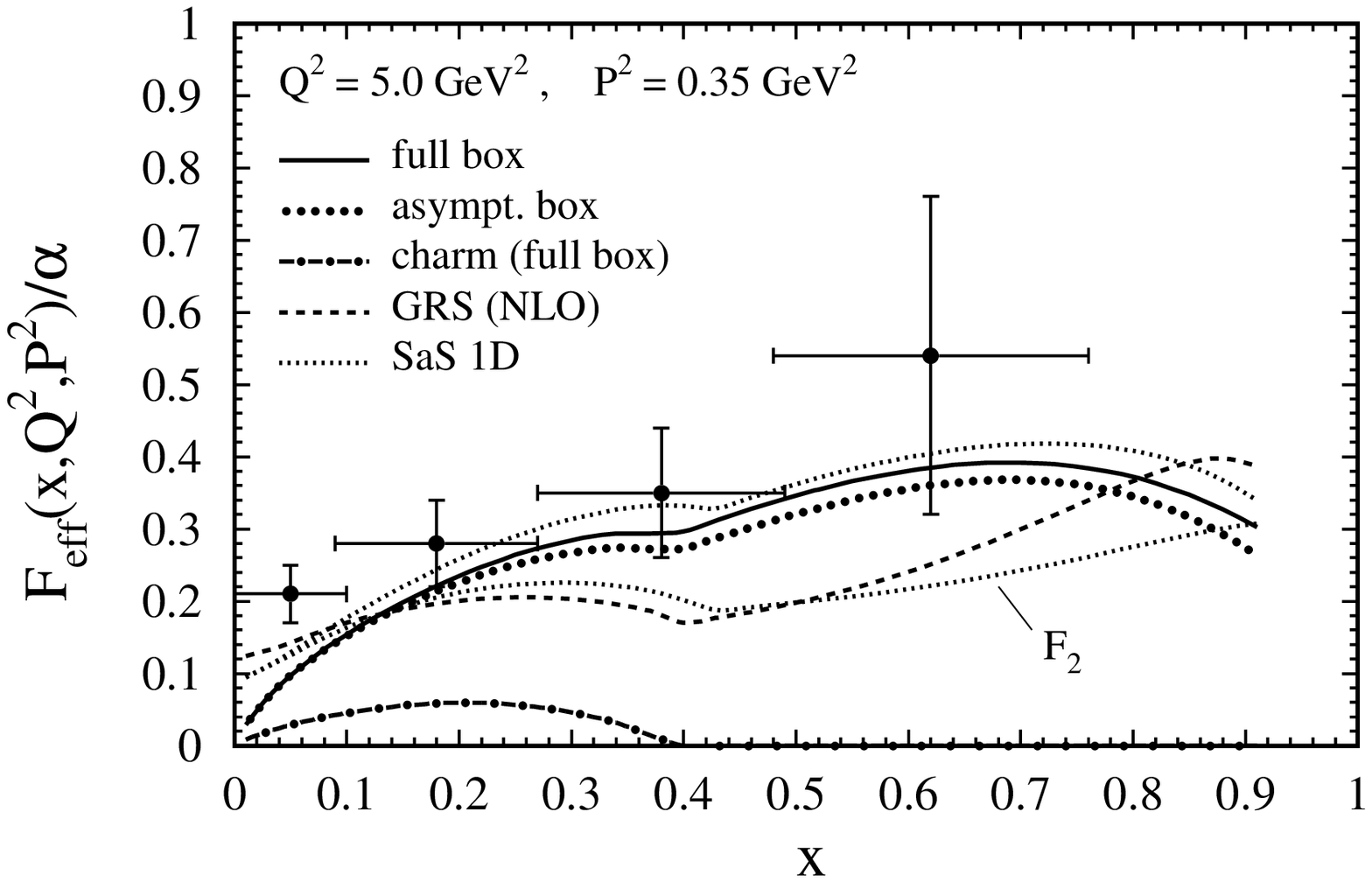,width=16cm}

\vspace*{0.3cm}
{\large\bf Fig. 1}
\end{figure}

\newpage
\begin{figure}
\centering
\vspace*{-1cm}
\epsfig{figure=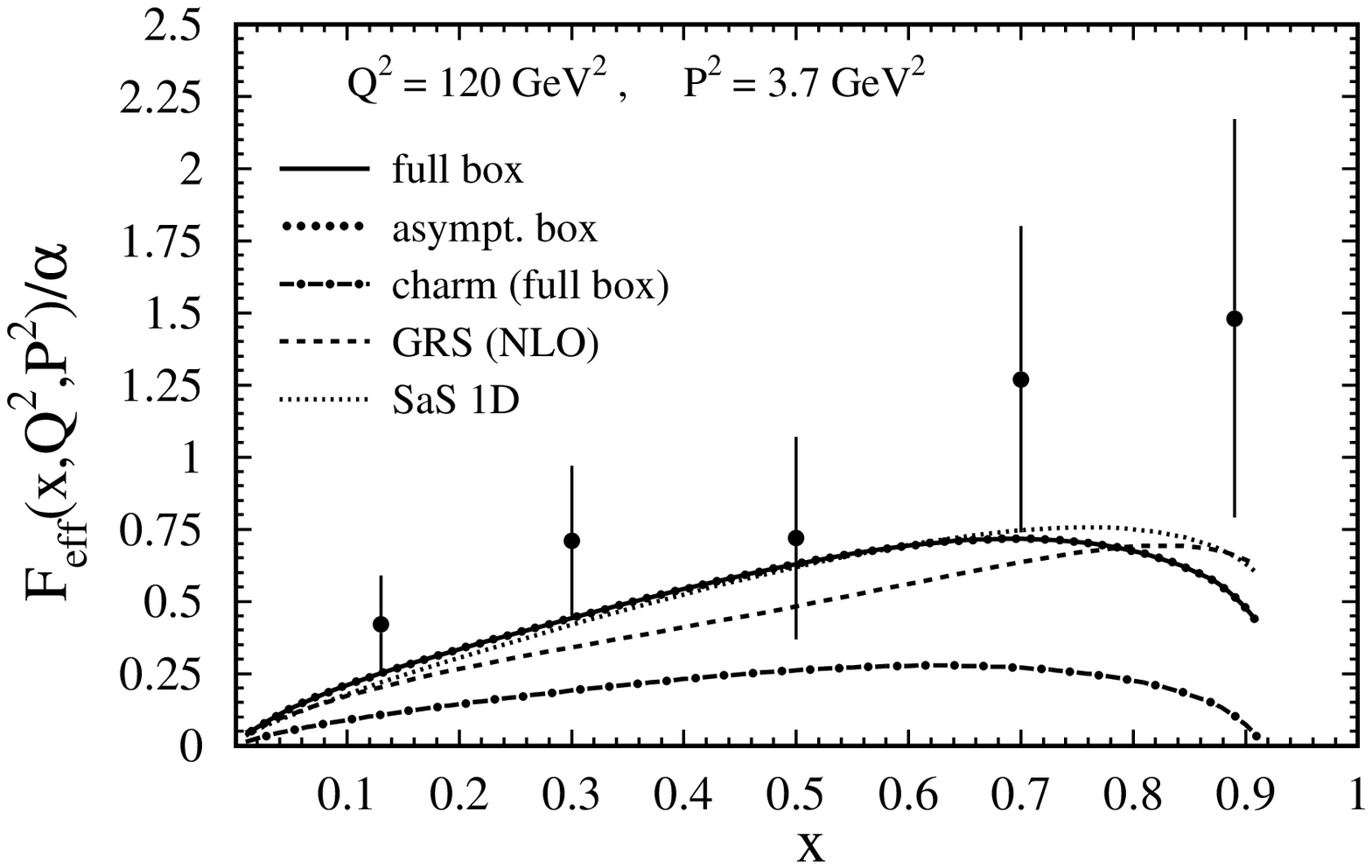,width=16cm}

\vspace*{0.3cm}
{\large\bf Fig. 2}
\end{figure}

\newpage
\begin{figure}
\centering
\vspace*{-1cm}
\epsfig{figure=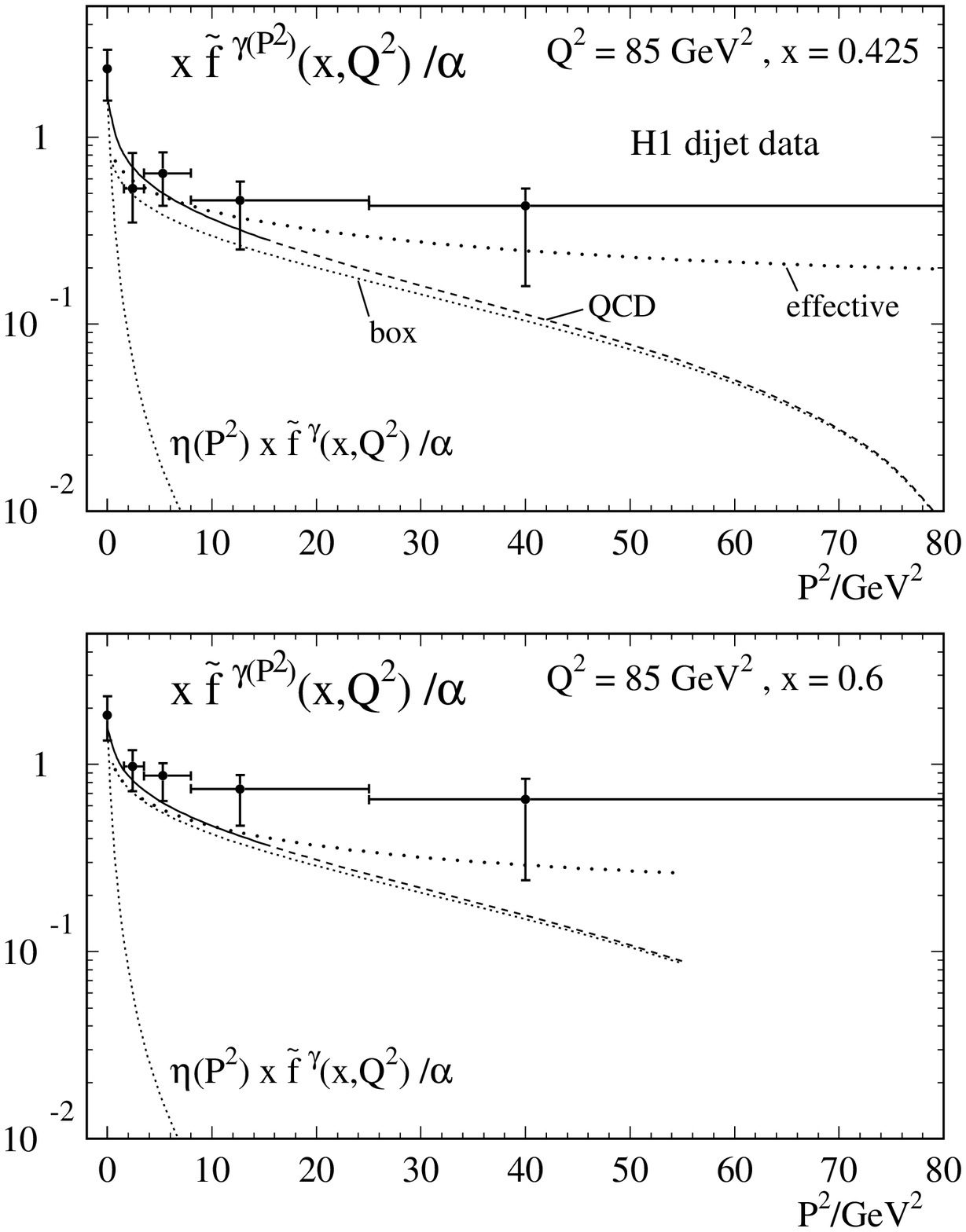,width=16cm}

\vspace*{0.3cm}
{\large\bf Fig. 3}
\end{figure}

\newpage
\begin{figure}
\centering
\vspace*{-1cm}
\epsfig{figure=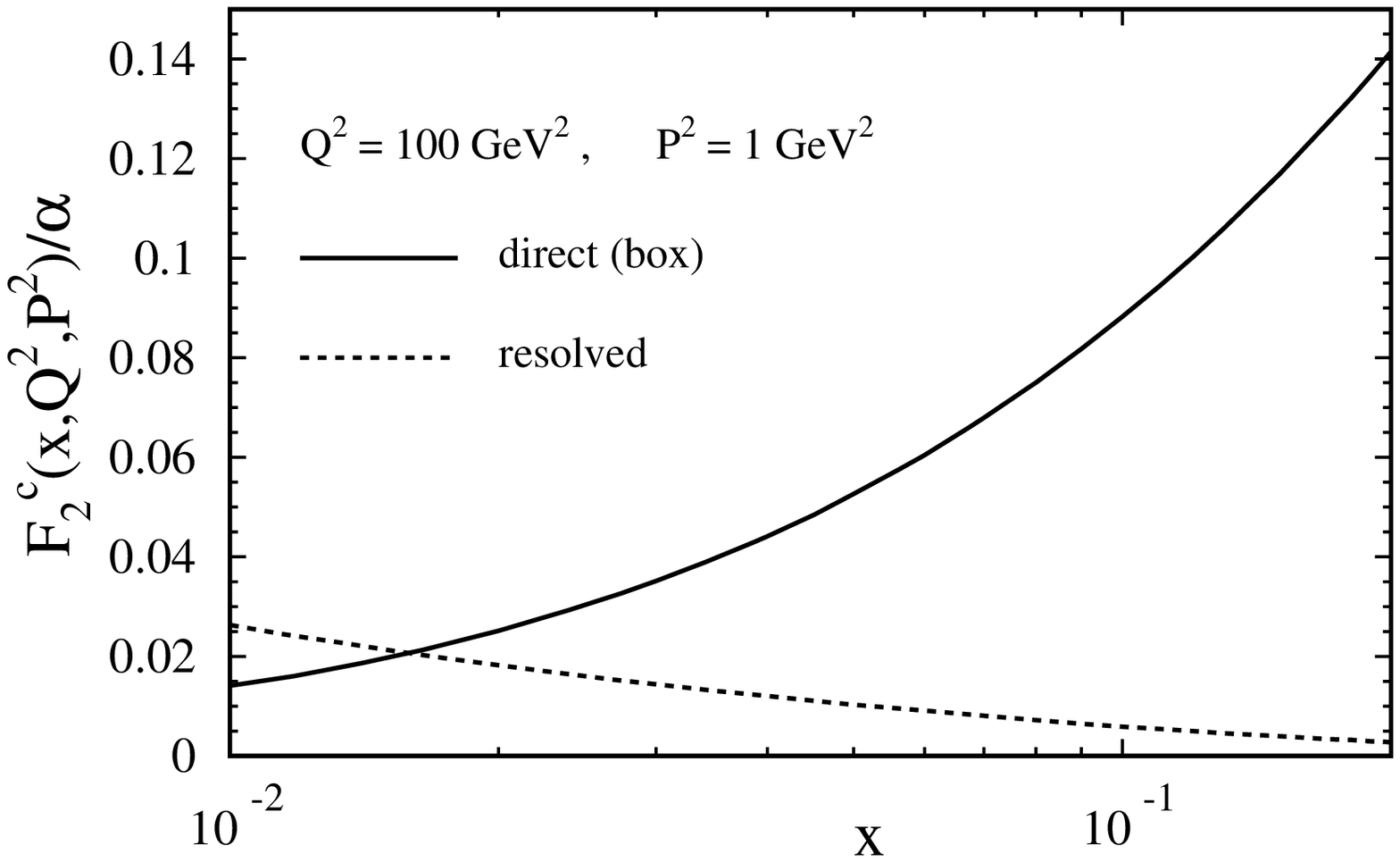,width=16cm}

\vspace*{0.3cm}
{\large\bf Fig. 4}
\end{figure}

\newpage
\begin{figure}
\centering
\vspace*{-1cm}
\epsfig{figure=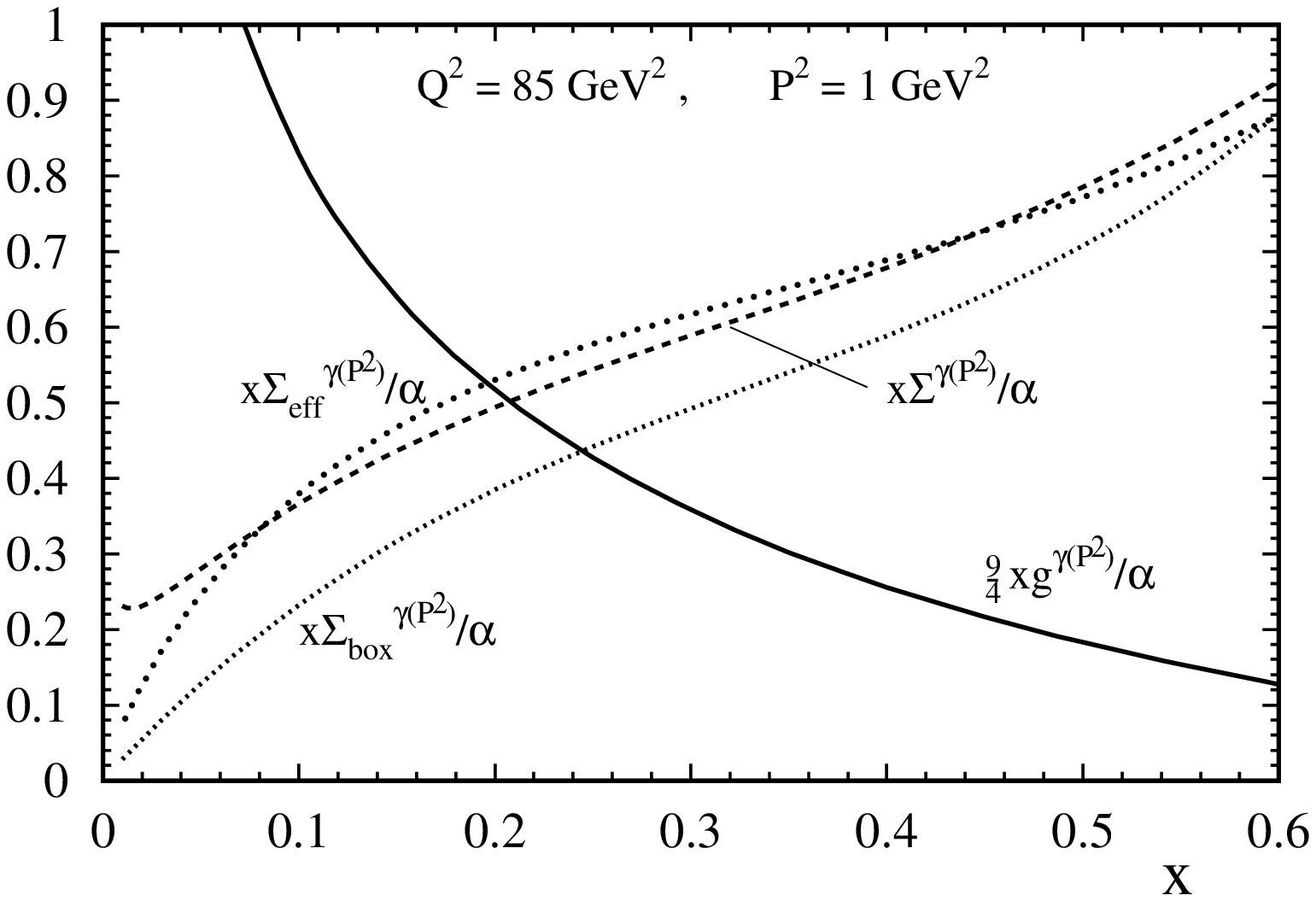,width=16cm}
\epsfig{figure=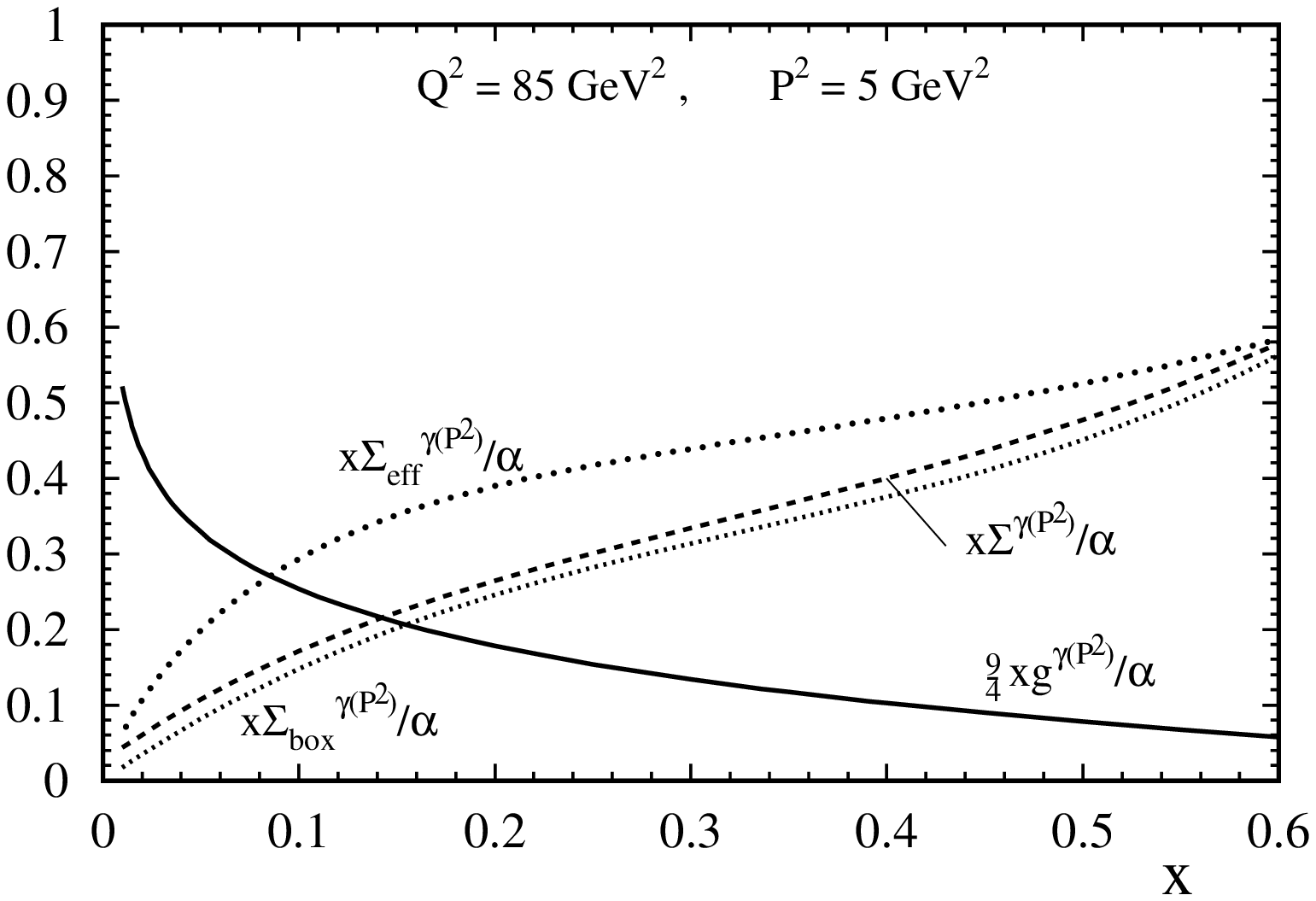,width=16cm}

\vspace*{0.3cm}
{\large\bf Fig. 5}
\end{figure}

\end{document}